# Structural Phase Stability in Fluorinated Calcium Hydride

R. Varunaa[1] and P. Ravindran[1, 2,*]

[1]Department of Physics, Central University of Tamil Nadu, Thiruvarur, Tamil Nadu, 610101, India
[2]Center of Material Science and Nanotechnology and Department of Chemistry, University of Oslo, Box 1033 Blindern, N-0315 Oslo, Norway    email: raviphy@cutn.ac.in

**Abstract.** In order to improve the hydrogen storage properties of calcium hydride ($CaH_2$), we have tuned its thermodynamical properties through fluorination. Using *ab-initio* total energy calculations based on density functional theory, the structural stability, electronic structure and chemical bonding of $CaH_{2-x}F_x$ systems are investigated. The phase transition of fluorinated systems from orthorhombic to cubic structure has been observed at 18% fluorine doped $CaH_2$. The phase stability analysis shows that $CaH_{2-x}F_x$ systems are highly stable and the stability is directly correlating with their ionicity. Density of states (DOS) plot reveals that $CaH_{2-x}F_x$ systems are insulators. Partial DOS and charge density analyses conclude that these systems are governed by ionic bonding. Our results show that H closer to F can be removed more easily than that far away from F and this is due to disproportionation induced in the bonding interaction by fluorination.

## INTRODUCTION

Calcium hydride ($CaH_2$) is considered as a saline hydride which has been used as hydrogen storage material due to its high hydrogen-to-metal ratio and also used for hydrogen production techniques. Experimentally observed that $CaH_2$ has high enthalpy of formation (-181.5 kJ/mol) and hence the dehydrogenation temperature is also high (>600 °C). The advantages of using $CaH_2$ as hydrogen storage materials due to relative abundance, cheap and high gravimetric density. Reducing the decomposition temperature of $CaH_2$ is the challenging task for onboard applications. Doping hydrogen with other isovalent element can improve the hydrogen storage properties of hydrides. Utke *et al.* [1] reported that metal-fluorine-catalyzed $Ca(BH_4)_2$ improves the kinetics of $Ca(BH_4)_2$ and also hydrogen storage capacity. In this paper we are reporting the structural phase stability and transition, electronic structure, chemical bonding, and H site energy for fluorinated $CaH_2$.

## STRUCTURAL ASPECTS AND COMPUTATIONAL DETAILS

The ground state crystal structure of $CaH_2$ and calcium fluoride ($CaF_2$) are shown in Fig. 1. At ambient condition, $CaH_2$ crystallize in the orthorhombic structure (Pnma) while $CaF_2$ crystallize in face centered cubic (Fm-3m). The unit cell of $CaH_2$ contains four calcium atoms and eight hydrogen atoms, where each Ca is surrounded by nine H atoms. Contrarily each hydrogen of type1 (H1) and type2 (H2) is surrounded by four and three Ca atoms, respectively. Similarly $CaF_2$ contains four calcium and eight fluorine atoms in its unit cell and each F is tetrahedrally coordinated with Ca.

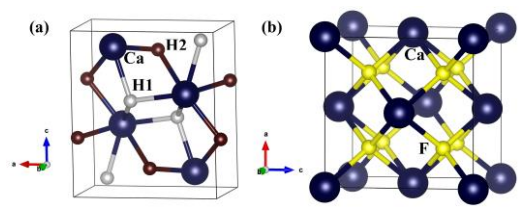

**FIGURE 1.** Crystal structure of (a) $CaH_2$ (Pnma) and (b) $CaF_2$ (Fm-3m).

The total energy calculations were done using the projector augmented plane wave method (PAW) implemented in Vienna abinitio simulation package (VASP). [2] The generalized gradient approximation proposed by Perdew, Burke and Ernzrhof (GGA-PBE) was used for structural optimizations (force as well as stress are minimized). Fluorination was achieved by using supercell approach. We used an energy cut off of

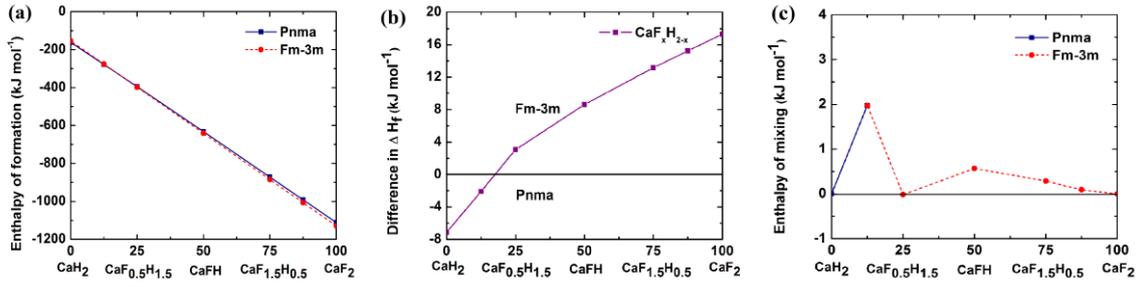

**FIGURE 2.** (a) Calculated enthalpy of formation, (b) difference in enthalpy of formation, and (c) calculated enthalpy of mixing as a function of fluorination.

300 eV for all the compositions considered here. The **k**-points were generated using the Monkhorst Pack method for structural optimization while the Gamma centered grid for electronic structure calculations.

## RESULTS AND DISCUSSION

The optimized and experimentally observed structural parameters of $CaH_2$ and $CaF_2$ are listed in Table 1 and the calculated equilibrium lattice parameters and volumes are in good agreement with the corresponding experimental data. The calculated enthalpy of formation ($\Delta H_f$) of $CaH_2$ and $CaF_2$ are found to be in good agreement with available experimental results in Table 2. From the $\Delta H_f$ values, it is to be concluded that $CaH_2$ prefer to be in orthorhombic (Pnma) while $CaF_2$ in cubic (Fm-3m) structure in consistent with experimental observations. The $\Delta H_f$ of $CaH_{2-x}F_x$ as a function of fluorine concentration is shown in Fig. 2(a). It is to be noted that all the fluorinated systems are relatively more stable compared with pure $CaH_2$. The difference in the enthalpy of formation between cubic and orthorhombic structure of $CaH_{2-x}F_x$ systems are also calculated and displayed in Fig. 2(b). From that we found the phase changing is occurring from Pnma to Fm-3m around 18% fluorine substitution in $CaH_2$.

**TABLE 1.** Theoretically calculated and experimentally observed (enclosed in bracket) structural parameters.

| Structure type | Lattice parameter (Å) | | | Volume (Å³) |
|---|---|---|---|---|
| | a | b | c | |
| $CaH_2$ (Pnma) | 5.8806 (5.9480 | 3.5619 3.6070 | 6.7537 6.8520) | 141.47 (147.01) [3] |
| $CaF_2$ (Fm-3m) | 5.4585 (5.4712 | 5.4585 5.4712 | 5.4585 5.4712) | 162.64 (163.78) [4] |

The enthalpy of mixing ($\Delta H_m$) of $CaH_{2-x}F_x$ systems in their ground state structure are calculated and plot between $\Delta H_m$ and composition is shown in Fig. 2(c). Interestingly the calculated $\Delta H_m$ of $CaH_{2-x}F_x$ systems are positive (except in case of 25% fluorination where the value is -15.6 J/mol) and very low implies that these compounds may expected to form at reasonable thermodynamic conditions. In support our finding, Pinatel *et al.* [5] discussed phase diagram of fluorinated $CaH_2$ by both experimentally and theoretically in their paper.

**TABLE 2.** Calculated enthalpy of formation per f.u.

| Structure type | Calculated $\Delta H_f$ (kJ/mol) | Experimentally observed $\Delta H_f$ (kJ/mol) |
|---|---|---|
| $CaH_2$ (Pnma) | -161.31 | -181.50 |
| $CaH_2$ (Fm-3m) | -154.19 | |
| $CaF_2$ (Pnma) | -1111.53 | |
| $CaF_2$ (Fm-3m) | -1128.81 | -1219.60 |

The chemical bonding between the constituents has been studied using partial density of states (DOS). The obtained bandgap of $CaH_2$ is 3.02 eV and it is very close to value 2.97 eV calculated by Yinwei *et al.*. [6] In Fig. 3(a), it is observed that the valence band of $CaH_2$ mainly originating from H states indicating the presence of ionic bonding between the constituents. For CaHF, the calculated bandgap is 2.71 eV. The F-$p$ states are well localized (-6 eV to -4 eV) and H-$s$ states are dominating the valence band (see Fig 3(b)).

Due to high electronegativity, F outermost states are completely filled resulting this well localized electronic states. As F draws more charges towards itself, it weakens the Ca-H bond as evident from our charge density analysis. This results reflects that phase stability of the system is governed mainly by ionic bonding. While seeing the pure fluoride system, the bandgap increased to 7.26 eV and in agreement with previous calculations. DOS analysis of $CaF_2$ shows (see Fig. 3(c)) that there is a strong ionic bonding is present between Ca and F.

The variation in bandgap as a function of fluorine substitution is shown in Fig. 4(a). The increase trend in bandgap with fluorine substitution is due to increase in ionicity. The calculated H site energy versus composition (see Fig. 4(b)) concludes that the energy required to remove H closer to F is low compared to H farer from F. So this indicating that F draws more charge towards itself due to its high electronegativity and hence it weakens the Ca-H bond closer to it.

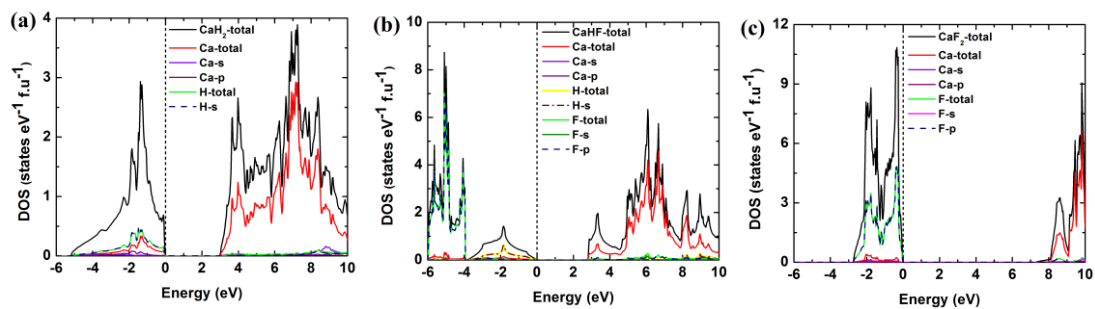

**FIGURE 3.** Total and partial DOS for (a) $CaH_2$ (Pnma), (b) CaHF (Fm-3m), and (c) $CaF_2$ (Fm-3m). Fermi level is set to zero.

Charge density plot of $CaH_2$ (see figure 5(a)) shows that there is an isotropic charge distribution is present in the constituents of $CaH_2$ indicating ionic nature of the compound. Fig. 5(b), shows that the charge density between Ca and H is reduced in CaHF compared with $CaH_2$ and also the charge density between Ca and F is large indicating that fluorination brings disproportionate bonding. Even though there is an isotropic charge distribution is seen in CaHF, still there is anisotropic charge distribution present at the H site indicating small amount of covalency is present in this system. For fluoride case (see Fig. 5(c)), it is clearly seen that $CaF_2$ has perfect ionic bonding due to an isotropic distribution of charges at both Ca and F sites.

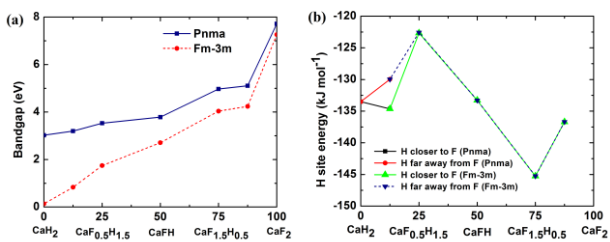

**FIGURE 4.** (a) Bandgap versus composition plot and (b) H site energy versus composition plot.

## CONCLUSIONS

The total energy calculations for fluorinated $CaH_2$ were done using VASP. The calculated enthalpy of formation reveals that the $CaH_{2-x}F_x$ systems are highly stable. The structural phase transition occurred at 18% fluorinated $CaH_2$. The phase mixing analysis shows that these systems can be formed at reasonable thermodynamical conditions. Partial DOS and charge density plots concludes that these systems are governed mainly by ionic bonding. From the calculated H site energy, we observed that one can easily remove hydrogen closer to fluorine compared to hydrogen far away from fluorine due to the formation of disproportionate bond. Also these systems are insulators and the bandgap increased from hydride to fluoride systems due to increase in ionicity.

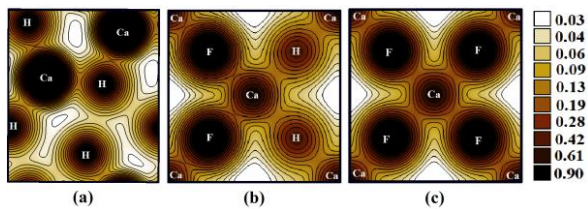

**FIGURE 5.** Charge density plot for (a) $CaH_2$ (Pnma), (b) CaHF (Fm-3m), and (c) $CaF_2$ (Fm-3m). Planes are chosen such a way that bond between the atoms are clearly seen.

## ACKNOWLEDGMENTS

The authors are grateful to the Department of Science and Technology, India for the funding support via Grant No. SR/NM/NS-1123/2013 and the Research Council of Norway for computing time on the Norwegian supercomputer facilities.